# Relational graph convolutional networks for predicting blood-brain barrier penetration of drug molecules

Yan Ding, Xiaoqian Jiang and Yejin Kim*

Center for Secure Artificial Intelligence for Healthcare, School of Biomedical Informatics
The University of Texas Health Science Center at Houston, Houston, TX, USA

*To whom correspondence should be addressed.

## Abstract

**Motivation:** Evaluating the blood-brain barrier (BBB) permeability of drug molecules is a critical step in brain drug development. Traditional methods for the evaluation require complicated *in vitro* or *in vivo* testing. Alternatively, *in silico* predictions based on machine learning have proved to be a cost-efficient way to complement the *in vitro* and *in vivo* methods. However, the performance of the established models has been limited by their incapability of dealing with the interactions between drugs and proteins, which play an important role in the mechanism behind the BBB penetrating behaviors. To address this limitation, we employed the relational graph convolutional network (RGCN) to handle the drug-protein interactions as well as the properties of each individual drug.

**Results:** The RGCN model achieved an overall accuracy of 0.872, an AUROC of 0.919 and an AUPRC of 0.838 for the testing dataset with the drug-protein interactions and the Mordred descriptors as the input. Introducing drug-drug similarity to connect structurally similar drugs in the data graph further improved the testing results, giving an overall accuracy of 0.876, an AUROC of 0.926 and an AUPRC of 0.865. In particular, the RGCN model was found to greatly outperform the LightGBM base model when evaluated with the drugs whose BBB penetration was dependent on drug-protein interactions. Our model is expected to provide high-confidence predictions of BBB permeability for drug prioritization in the experimental screening of BBB-penetrating drugs.

**Availability and Implementation:** The data and the codes are freely available at https://github.com/dingyan20/BBB-Penetration-Prediction.

**Contact:** yejin.kim@uth.tmc.edu

## 1. Introduction

Brain targeting drug development has long been suffering from poor penetration of drug molecules across the blood-brain barrier (BBB) (Terstappen *et al.*, 2021; Cecchelli *et al.*, 2007; Patel and Patel, 2017). The BBB, mainly formed by a monolayer of tightly packed endothelial cells, is a physical, metabolic and transport barrier that strictly regulates the transfer of substances between the blood and the neural tissues (Segarra *et al.*, 2021; Banks, 2016; Abbott *et al.*, 2010). Due to this barrier, most drug molecules cannot enter the brain, thus showing no therapeutic effects on central nervous system (CNS) disorders. Therefore, evaluating the BBB permeability of brain targeting drug candidates has become a critical step in the development of therapeutic agents for CNS diseases.

Conventionally, the BBB permeability of drug molecules has been evaluated with *in vitro* BBB models (Bagchi *et al.*, 2019; Sivandzade and Cucullo, 2018) or directly by animal experiments (Bicker *et al.*, 2014; Palmer and Alavijeh, 2013). Those methods are expensive, time-consuming, labor-intensive and unscalable. In contrast, *in silico* modeling based on machine learning is able to make predictions with good accuracy while largely avoiding the drawbacks of the conventional methods (Vatansever *et al.*, 2021).

Previously, different types of *in silico* models have been developed for predicting the BBB permeability (Shi *et al.*, 2021; Singh *et al.*, 2020; Alsenan *et al.*, 2020; Wang *et al.*, 2018; Garg and Verma, 2006; Li *et al.*, 2005), including logistic regression (Plisson and Piggott, 2019), support vector machine (SVM) (Yuan *et al.*, 2018), decision tree (Andres and Hutter, 2006), random forest (Martins *et al.*, 2012) and light gradient boosting machine (LightGBM) (Shaker *et al.*, 2021). Despite the variety of the algorithms involved, the input features for training and prediction have mostly been limited to the structures or the properties of the drug molecules. The most commonly used features are the physicochemical properties, such as the molecular weight, the polar surface area, and the octanol/water partitioning coefficient (LogP) (Pajouhesh and Lenz, 2005; Zhang *et al.*, 2008). In particular, Xu et al. introduced a special set of drug features, which combined clinical phenotypes (side effects and indications) of a drug molecule with its physicochemical properties (Gao *et al.*, 2017; Miao *et al.*, 2019). Additionally, the probability of being a substrate of efflux transporters has also been used as a drug feature in the prediction of BBB penetration (Dolghih and Jacobson, 2013; Lingineni *et al.*, 2017; Garg *et al.*, 2015).

The previously established models only take the features of each individual drug as the input. However, the BBB penetration of a drug molecule depends not only on the properties of the drug itself, but can also be affected by drug-protein interactions (Vatansever *et al.*, 2021; Yuan *et al.*, 2018), which play an important role in the mechanisms behind BBB penetration (Figure 1). In the case where a lipophilic drug crosses the BBB through passive diffusion, the BBB permeability is mainly determined by the physicochemical properties of the drug (Alavijeh *et al.*, 2005; Pajouhesh and Lenz, 2005). However, drugs can also be transferred across the BBB by the influx transporters on the surface of the endothelial cells (Sanchez-Covarrubias *et al.*, 2014; Pardridge, 2012). For example, levodopa (Whitfield *et al.*, 2014) and melphalan (Cornford *et al.*, 1992) enter the brain through neutral amino acid transporters. In contrast, active efflux transporters such as P-glycoprotein (P-gp) and multidrug resistance-associated protein 1 (MRP1) can pump drugs out of the endothelial cells and thus expel the drugs back to the blood (Cecchelli *et al.*, 2007; Qosa *et al.*, 2015). In addition, protein binding can also influence the BBB permeability of drugs. Drugs bound to plasma proteins such as Albumin tends to stay in the blood because the proteins can hardly cross the BBB (Wanat, 2020; Bohnert and Gan, 2013); only the unbound fraction of the drugs in the plasma may have a chance to go across the BBB. Besides, structural similarity between drug molecules can also be exploited to infer the BBB permeability of drugs. Structurally similar drug molecules are likely to show similar physicochemical properties and may bind to the same proteins (Muegge and Mukherjee, 2016; Eckert and Bajorath, 2007; Martin *et al.*, 2002). Thus, the drug-drug similarity is closely related to the factors affecting BBB penetration. Altogether, the drug-protein interactions and the drug-drug similarity are important relations that can contribute to the prediction of BBB penetration. A model that can handle those relations, as well as the features of each drug molecule, is highly desirable.

Graph convolutional networks (GCNs) have been drawing more and more attention during the past few years (Kipf and Welling, 2017; Zitnik *et al.*, 2018; Kearnes *et al.*, 2016; Long *et al.*, 2020; Cai *et al.*, 2020; Baldassarre *et al.*, 2021; Schulte-Sasse *et al.*, 2021; Yuan and Bar-Joseph, 2020). In a GCN model, the data is presented in the form of a graph, with the samples as the nodes and the relations between the samples as the edges. The graph convolution operation updates the representation of each node through the aggregation of the messages passed from its neighboring nodes. Thereby, every node learns about its topological context while preserving its own features. Since graphs can efficiently represent the different types of interactions among drugs and proteins, the GCN model emerges as a promising model for the prediction of BBB permeability. The vanilla GCN treats all relations in a graph as the same. However, different drug-protein relations can have different effects on BBB penetration. For example, drugs related to the influx transporters are expected to be BBB penetrating, whereas drugs binding to plasma proteins are unlikely to cross the BBB. As a result, the vanilla GCN model is not suitable for predicting BBB permeability because it does not distinguish between different types of relations. To solve that problem, a variant of the GCN, which is called relational GCN (RGCN), can be employed since it is able to handle heterogeneous graphs (i.e., graphs containing various types of nodes and edges) by assigning different weight matrices to different kinds of relations (Schlichtkrull *et al.*, 2018; Thanapalasingam *et al.*, 2021; Shu *et al.*, 2021).

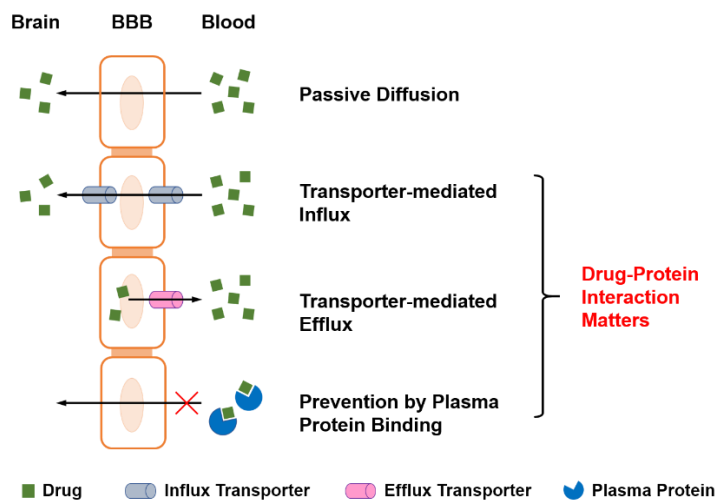

Figure 1. Mechanisms of BBB penetration for small-molecule drugs.

Herein, we developed an RGCN model for predicting the BBB permeability of drug molecules (Figure 2). The model was trained with a graph where the nodes represented drugs and proteins and the edges represented drug-protein/drug-drug relations. Initially, the training was conducted with the drug-protein interactions and the Mordred descriptors (Moriwaki *et al.*, 2018) of the drug molecules as the input. The resultant model showed better performance than most of the previously reported ones (Shaker *et al.*, 2021; Alsenan *et al.*, 2021), highlighting the important role of the drug-protein interactions in predicting BBB permeability. Moreover, the drug-drug similarity was integrated into the RGCN model to further improve the predictive capability of the model. Furthermore, we also demonstrated that the RGCN was much more powerful than other models in classifying the drugs whose BBB penetration depended on drug-protein interactions.

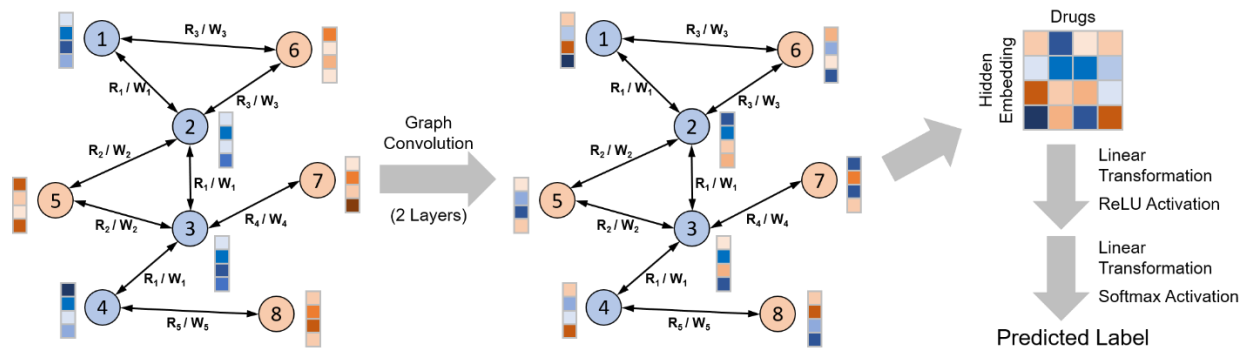

Figure 2. Schematic illustration of the RGCN model. First, the data is structured into a graph. Node 1-4 (blue) represent drugs and Node 5-7 (orange) stand for proteins. All the drug nodes are labeled according to the BBB permeability of the drug. Next to the nodes are the node features. The Mordred descriptors are used as the features of the drug nodes. The features of the protein nodes are treated as learnable parameters. $R_1$, $R_2$, $R_3$, $R_4$ and $R_5$ represent the five types of relations between the nodes, i.e., the drug-drug similarity, Drug-Protein Influx, Drug-Protein Efflux, Drug-Protein Carrier and Drug-Protein Other. $W_1$, $W_2$, $W_3$, $W_4$ and $W_5$ are the corresponding matrices used for the relation-specific transformation of the feature vectors of the neighboring nodes. The hidden embeddings of the drugs are obtained through graph convolution and are subsequently passed through a classifier. Two convolution layers have been used in our experiments. The classifier is composed of two linear layers, which are followed by a ReLU activation layer and a softmax activation layer, respectively.

## 2. Material and methods

### 2.1 Data collection

Drugs labeled with their BBB permeability were collected from the literature (Martins *et al.*, 2012; Shaker *et al.*, 2021; Alsenan *et al.*, 2021; Wang *et al.*, 2018). The Simplified Molecular-Input Line-Entry System (SMILES) strings were used to identify the drugs. The drugs with invalid SMILES strings were removed. The duplicates were also deleted. The drug-protein interactions were gathered from STITCH (Szklarczyk *et al.*, 2016). The interactions with a combined score of 500 or higher were used. We further classified the drug-protein interactions into four categories according to the function of the involved protein. The proteins belonging to the solute carrier (SLC) family usually function as transporters facilitating the uptake of drugs into the brain (Nałęcz, 2017). The drug-protein interactions involving the SLC family were grouped into one category, which was named Drug-Protein Influx. The ATP-binding cassette (ABC) family contains efflux transporters that inhibit the penetration of drugs across the BBB (Morris *et al.*, 2017). And the interactions between the drugs and the ABC transporters were labeled as Drug-Protein Efflux. The carrier proteins bind to certain drugs and modify the pharmacokinetics of the drugs (e.g., BBB penetration) (Wishart *et al.*, 2018). The interactions involving the carrier proteins constituted the third category, Drug-Protein Carrier. All the other drug-protein interactions fell into the fourth category called Drug-Protein Other. The four types of drug-protein interactions would be handled as four different kinds of relations in the RGCN model.

## 2.2 Drug features

Mordred descriptors (Moriwaki *et al.*, 2018) were used as the drug features. The 2D Mordred descriptors were generated using the molecular featurizer provided by the DeepChem library (https://deepchem.readthedocs.io/en/latest/api_reference/featurizers.html). The dimension of the Mordred descriptors was 1613. The Mordred descriptors were standardized before they were fed to the RGCN model as the node features for the drugs.

## 2.3 Drug-drug similarity

The drug-drug similarity was evaluated by the Tanimoto score, which was calculated with the 2D pharmacophore fingerprints of the drug molecules (Muegge and Mukherjee, 2016). The pharmacophore fingerprints were generated using the cheminformatics toolkit RDKit (https://github.com/rdkit/rdkit). The definitions of chemical features were from the file BaseFeatures provided by RDKit. The chemical feature called ZnBinder was skipped. The distance bins were set to (0, 2), (2, 4), (4, 6), and (6, 10). Two-point and three-point pharmacophores were identified. And the fingerprints were generated accordingly. The threshold of the similarity score was optimized as a hyperparameter.

## 2.4 RGCN model

The RGCN is an extension of the classic GCN and shares the same essential algorithm with the GCN (Schlichtkrull *et al.*, 2018; Thanapalasingam *et al.*, 2021). In the GCN model, all the feature vectors of the neighboring nodes are transformed with the same weight matrix. In contrast, for the RGCN, the concept of relation types has been introduced. Herein, the vector transformation is relation-specific, i.e., each type of relation is assigned a distinct weight matrix. As a result, the RGCN model is able to handle heterogeneous graphs.

In the present work, an RGCN model was built using Pytorch Geometric (Fey and Lenssen, 2019). The RGCNConv module provided by Pytorch Geometric was directly used as the convolution layer in our model. There were two convolution layers followed by two linear layers (Figure 2). The dimensions of the output feature vectors of the first three layers, as well as the dropout rate, were optimized as hyperparameters using Optuna (Akiba *et al.*, 2019), which is an automatic hyperparameter optimization framework. The features of the gene nodes were treated as part of the model parameters and would be updated during training.

## 2.5 Model training and evaluation

The Adam optimizer was used for training because of its consistent high efficiency (Choi *et al.*, 2019). Two losses, i.e., the negative log-likelihood loss (NLLLoss) and the triplet margin loss (Balntas *et al.*, 2016), were calculated. The NLLLoss was used as the classification loss and was calculated with the final output of the model. The triplet margin loss was calculated with the hidden drug embeddings (the output of the first linear layer of the model) using PyTorch Metric Learning (Musgrave *et al.*, 2020). Minimizing the triplet margin loss could force the hidden embeddings of the drugs with the same label to cluster, which would facilitate the classification. To handle the two losses, we adopted the method called projecting conflicting gradients (PCGrad) (Yu *et al.*, 2020), which could mitigate the potential conflict between the gradients generated from the two loss functions.

The data collected were structured into a graph. The drugs were split into the training dataset, the validation dataset and the testing dataset at a ratio of 70:15:15 in a stratified manner. 10-fold cross-validation was first conducted with the training dataset for hyperparameter tuning. The hyperparameters were tuned using Optuna. The mean values of the validation scores obtained from the 10-fold cross-validation were used to compare different sets of hyperparameters. The hyperparameters, as well as the search space for each hyperparameter, are listed in the supporting information. Then, the model was trained with the training dataset and evaluated with the validation dataset. The area under the receiver operating characteristic (AUROC) for the validation dataset was monitored for early stopping during the training. Specifically, the training would be halted if the AUROC for the validation dataset stopped increasing. Finally, the trained model was evaluated with the testing dataset.

The performance of the model was assessed by multiple metrics, including the accuracy, the sensitivity (the true positive rate), the specificity (the true negative rate), the Matthews correlation coefficient (MCC), the AUROC, and the area under the precision-recall curve (AUPRC). The AUPRC was obtained by computing the average precision. Particularly, the average precision was calculated with the nonpenetrating drugs as the positive class, because the nonpenetrating drugs were the minor group. The baseline value of the AUPRC was 0.297, which was obtained by dividing the number of the nonpenetrating drugs (1178) by the total number of the drugs in the dataset (3961), given that the data was split in a stratified manner.

## 3. Results

### 3.1 RGCN model featuring the drug-protein interactions

An RGCN model was first developed to demonstrate the role of drug-protein interactions in the prediction of BBB permeability.

A graph was prepared with 3961 drugs and 4264 relevant proteins as the nodes. All the drug nodes were labeled either as penetrating drugs (2783) or nonpenetrating drugs (1178). Edges were established according to the drug-protein interactions. There were 25021 drug-protein edges, among which 1007 were for Drug-Protein Influx, 308 were for Drug-Protein Efflux, 417 were for Drug-Protein Carrier, and 23289 were for Drug-Protein Other. The Mordred descriptors, which represented the structural and the physicochemical properties of the drug molecules, were introduced as the node features for the drugs. The node features for the proteins were treated as learnable parameters.

The RGCN model was trained and evaluated with the prepared graph (Section 2.5). The results from both the 10-fold cross-validation and the testing are shown in Table 1. Specifically, the model gave an overall accuracy of 0.872 for the testing dataset, with an AUROC of 0.919. The sensitivity and the specificity were 0.919 and 0.763, respectively. Since the dataset was imbalanced, two more metrics, the AUPRC and the MCC were also calculated to further evaluate the model. The AUPRC was 0.838, which was remarkably higher than the baseline value 0.297 (Section 2.5). And a MCC of 0.691 was obtained for the testing dataset, indicating fairly strong correlation between the drugs' true labels and the predicted ones. The performance of the RGCN is among the best results reported in recent papers.

Meanwhile, we employed the LightGBM as the base model for comparison because it is one of the most powerful models for tabular data. Shaker et al. lately reported a LightGBM model that showed top performance in predicting BBB permeability (Shaker *et al.*, 2021). However, the high performance was achieved partially due to data leakage, which was caused by the duplicates in their dataset. Canonical SMILES strings instead of the raw SMILES strings should have been used to identify the drugs. Therefore, we trained a LightGBM model ourselves using the Mordred descriptors as the input. And the model was rigorously optimized using Optuna (Akiba *et al.*, 2019), which is an automatic hyperparameter optimization framework. As shown in Table 1, the RGCN model was better than the LightGBM model in terms of the accuracy, the MCC and the AUROC. It is also worth noting that the RGCN model gave a much higher specificity while offering a lower sensitivity (0.919 for the testing dataset) that was still quite high. That means the RGCN model was good at predicting the labels for both the positive class and the negative class. In contrast, the LightGBM model performed really well in classifying the positive class, but did a poor job with the negative class. The results above indicate that the RGCN model is overall better than the LightGBM model, although the difference is not very big. The notably higher MCC, along with the more balanced predictive ability, results from the capability of the RGCN to learn the topological structure of the graph, i.e., the drug-protein interactions.

In addition, a support vector machine (SVM) model was also trained with the same drug features, i.e., the Mordred descriptors. The SVM did not perform as well as the RGCN model, although it achieved the highest AUPRC for the testing dataset among the three models.

Table 1. Performance of the RGCN model with only the drug-protein interactions as the edges.

| Model | Experiment | Accuracy | Sensitivity | Specificity | MCC | AUROC | AUPRC |
|---|---|---|---|---|---|---|---|
| RGCN | Cross-Validation | 0.860 | 0.917 | 0.727 | 0.662 | 0.913 | 0.835 |
| | Testing | 0.872 | 0.919 | 0.763 | 0.691 | 0.919 | 0.838 |
| LightGBM | Cross-Validation | 0.860 | 0.953 | 0.641 | 0.653 | 0.918 | 0.847 |
| | Testing | 0.861 | 0.959 | 0.627 | 0.652 | 0.907 | 0.840 |
| SVM | Cross-Validation | 0.852 | 0.935 | 0.654 | 0.632 | 0.894 | 0.821 |
| | Testing | 0.857 | 0.935 | 0.672 | 0.646 | 0.909 | 0.854 |

**3.2 Drug-drug similarity boosts the performance of the RGCN model**

Drugs with similar structures tend to show similar properties and thus have similar BBB permeability. Introduction of drug-drug similarity to the RGCN can enhance the networking between the drug nodes and facilitate the message passing within the graph, leading to better performance of the model.

Herein, the drug-drug similarity was added to the graph as the fifth type of node relation. Edges were established between two drugs if the similarity score (Section 2.3) was higher than the similarity threshold. The similarity threshold determined the number of the drug-drug edges. And it was optimized as a hyperparameter. Table 2 shows the performance of the RGCN model with the similarity threshold set to different values. The best scores were obtained when the threshold was set to 0.9 and 0.6. Considering the number of edges that would be added, 0.9 was chosen as the threshold just to make the model simpler. In this case, 5620 drug-drug edges were integrated into the data graph. The drug-drug similarity was included in all the further experiments with the similarity threshold kept at 0.9.

Table 2. Optimization of the similarity threshold.

| Similarity Threshold | Number of Edges (Drug-Drug Similarity) | Accuracy | MCC | AUROC | AUPRC |
|---|---|---|---|---|---|
| 0.9 | 5620 | 0.876 | 0.695 | 0.926 | 0.865 |
| 0.8 | 8834 | 0.876 | 0.694 | 0.917 | 0.846 |
| 0.7 | 15167 | 0.850 | 0.655 | 0.922 | 0.864 |
| 0.6 | 25770 | 0.876 | 0.695 | 0.925 | 0.868 |
| 0.5 | 48119 | 0.850 | 0.649 | 0.917 | 0.851 |

The performance of the RGCN model was boosted after drug-drug similarity was introduced. With the similarity threshold set to 0.9, the AUROC and the AUPRC scores for the testing dataset went up to 0.926 and 0.865, respectively (Table 2), which were now notably higher than the values given by the LightGBM model (0.907 and 0.840, Table 1) and the SVM model (0.909 and 0.854, Table 1).

### 3.3 The RGCN prevails when the drug-protein interaction matters

The RGCN gains advantage over the previous models by handling the drug-protein interactions. Thus, it is more effective in making predictions for the drugs whose BBB permeability depends on the drug-protein interactions. However, a large portion of the drugs in the testing dataset cross the BBB through passive diffusion with no drug-protein interaction involved. Out of the 595 drugs in the testing dataset, 184 were interacting with at least one protein in the data graph while 411 were not connected to any protein. To fully demonstrate the power of the RGCN model, we evaluated the RGCN model using the subsets of the testing dataset. Each subset contained the drugs interacting with a specific type of the protein in our data graph, as well as the drugs not interacting with the proteins.

Table 3. Performance of the RGCN model when it was evaluated with the subsets of the testing dataset. Note that one drug could interact with different proteins. Thus, there might be overlap between the subsets.

| Subset of Testing Dataset | Number of Drugs | Model | Accuracy | MCC | AUROC | AUPRC |
|---|---|---|---|---|---|---|
| Drugs interacting with the ABC protein family | 38 | RGCN | 0.737 | 0.453 | 0.832 | 0.902 |
|  |  | LightGBM | 0.553 | 0.309 | 0.800 | 0.901 |
| Drugs interacting with the SLC protein family | 45 | RGCN | 0.800 | 0.594 | 0.863 | 0.841 |
|  |  | LightGBM | 0.711 | 0.428 | 0.821 | 0.821 |
| Drugs interacting with the carrier proteins | 28 | RGCN | 0.786 | 0.579 | 0.810 | 0.785 |
|  |  | LightGBM | 0.571 | 0.141 | 0.721 | 0.709 |
| Drugs interacting with the other types of proteins | 177 | RGCN | 0.825 | 0.604 | 0.874 | 0.799 |
|  |  | LightGBM | 0.746 | 0.411 | 0.835 | 0.751 |
| All protein-interacting drugs | 184 | RGCN | 0.810 | 0.568 | 0.873 | 0.776 |
|  |  | LightGBM | 0.755 | 0.433 | 0.843 | 0.764 |
| Drugs not interacting with any protein | 411 | RGCN | 0.905 | 0.755 | 0.943 | 0.886 |
|  |  | LightGBM | 0.908 | 0.761 | 0.945 | 0.890 |

The results are shown in Table 4. As expected, the RGCN model clearly outperformed the LightGBM model for the protein-interacting drugs. For example, when evaluated with the SLC protein-interacting drugs in the test dataset, the RGCN model gave markedly higher accuracy, MCC, AUROC and AUPRC scores than the LightGBM model. Similar results were observed with the drugs interacting with the other types of the proteins. On the other hand, when the RGCN was tested with the drugs not connected with the proteins, the difference in performance between the RGCN and the LightGBM model became marginal. Therefore, when the RGCN was evaluated using the whole testing dataset, which contained more non-protein-interacting drugs (411) than protein-interacting drugs (184), the advantage of the RGCN was diminished. In other words, the performance of our model was, to a great extent, limited by the dataset we were using.

## 4. Discussion

Data availability remains a major challenge in predicting the BBB penetration of drug molecules. Among the drugs available, most cross the BBB through passive diffusion with no drug-protein interaction involved, thus limiting the performance of our RGCN model. Data augmentation serves as an efficient way to increase the amount of data. Particularly, Zhao et al. recently published a paper regarding the data augmentation for graph neural networks (Zhao *et al.*, 2020), which can potentially help with our current research. We believe the RGCN model will further manifest its power if more data, especially the drug-protein interactions involved in BBB penetration, can be generated through data augmentation.

The datasets used for predicting BBB penetration are usually imbalanced. Therefore, the AUPRC is a more appropriate metric for the model evaluation. In the present work, the model was trained by minimizing the NLLLoss and the triplet margin loss. And the AUPRC was calculated with the predicted probabilities and the true labels. Recently, Yang et al. reported a principled technical method to optimize AUPRC for deep learning (Qi *et al.*, 2021). With this approach, the authors were able to train their models by directly maximizing the AUPRC and achieve state-of-art performance. We consider the direct AUPRC optimization to be a promising strategy to enhance the prediction of BBB penetration and will employ it in our future work.

Currently, most of the drugs being studied in the field of BBB penetration are supposed to cross the BBB through the mechanism of passive diffusion. When dealing this type of drug, the RGCN model only showed minor improvement over the LightGBM model. However, it was much more time consuming due to its higher computational complexity. The LightGBM is still a good choice when drug-protein interaction is not involved in the BBB penetration of the drugs.

## 5. Conclusion

We have successfully established an RGCN model for the prediction of BBB permeability. In our study, different mechanisms of BBB penetration for small-molecule drugs were considered. For each mechanism, the factors that could affect the BBB permeability were identified. And the RGCN model was chosen to handle all the factors involved, i.e., the drug properties, the drug-protein interactions, and the drug-drug similarity. Therefore, our model can make predictions for small-molecule drugs penetrating the BBB through all kinds of mechanisms, while the previously built models only apply to lipophilic drugs that penetrate the BBB through passive diffusion. The

RGCN model represents a new strategy for predicting BBB permeability and can potentially serve as a generic platform for the prediction of other pharmacological properties of drug molecules.

## Acknowledgements

This work was supported by the National Institutes of Health [R01AG066749 to X.J. and Y.K., U01TR002062 to X.J.]; the Cancer Prevention & Research Institute of Texas [RR180012]; The University of Texas under the STARs program and the UTHealth startup program; and the Christopher Sarofim Family Professorship.

# References

Abbott,N.J. *et al.* (2010) Structure and function of the blood–brain barrier. *Neurobiol. Dis.*, **37**, 13–25.

Akiba,T. *et al.* (2019) Optuna: A Next-Generation Hyperparameter Optimization Framework. In, *Proceedings of the 25th ACM SIGKDD International Conference on Knowledge Discovery & Data Mining*, KDD '19. Association for Computing Machinery, New York, NY, USA, pp. 2623–2631.

Alavijeh,M.S. *et al.* (2005) Drug metabolism and pharmacokinetics, the blood-brain barrier, and central nervous system drug discovery. *NeuroRx*, **2**, 554–571.

Alsenan,S. *et al.* (2021) A deep learning approach to predict blood-brain barrier permeability. *PeerJ Comput. Sci.*, **7**, e515.

Alsenan,S. *et al.* (2020) A Recurrent Neural Network model to predict blood–brain barrier permeability. *Comput. Biol. Chem.*, **89**, 107377.

Andres,C. and Hutter,M.C. (2006) CNS permeability of drugs predicted by a decision tree. *QSAR Comb. Sci.*, **25**, 305–309.

Bagchi,S. *et al.* (2019) In-vitro blood-brain barrier models for drug screening and permeation studies: An overview. *Drug Des. Devel. Ther.*, **13**, 3591–3605.

Baldassarre,F. *et al.* (2021) GraphQA: protein model quality assessment using graph convolutional networks. *Bioinformatics*, **37**, 360–366.

Balntas,V. *et al.* (2016) Learning local feature descriptors with triplets and shallow convolutional neural networks. In, *Procedings of the British Machine Vision Conference 2016*. British Machine Vision Association, pp. 119.1-119.11.

Banks,W.A. (2016) From blood–brain barrier to blood–brain interface: new opportunities for CNS drug delivery. *Nat. Rev. Drug Discov.*, **15**, 275–292.

Bicker,J. *et al.* (2014) Blood–brain barrier models and their relevance for a successful development of CNS drug delivery systems: A review. *Eur. J. Pharm. Biopharm.*, **87**, 409–432.

Bohnert,T. and Gan,L.-S. (2013) Plasma protein binding: From discovery to development. *J. Pharm. Sci.*, **102**, 2953–2994.

Cai,R. *et al.* (2020) Dual-dropout graph convolutional network for predicting synthetic lethality in human cancers. *Bioinformatics*, **36**, 4458–4465.

Cecchelli,R. *et al.* (2007) Modelling of the blood–brain barrier in drug discovery and development. *Nat. Rev. Drug Discov.*, **6**, 650–661.

Choi,D. *et al.* (2019) On Empirical Comparisons of Optimizers for Deep Learning. arXiv:1910.05446.

Cornford,E.M. *et al.* (1992) Melphalan penetration of the blood-brain barrier via the neutral amino acid transporter in tumor-bearing brain. *Cancer Res.*, **52**, 138–43.

Dolghih,E. and Jacobson,M.P. (2013) Predicting Efflux Ratios and Blood-Brain Barrier Penetration from Chemical Structure: Combining Passive Permeability with Active Efflux by P-Glycoprotein. *ACS Chem. Neurosci.*, **4**, 361–367.

Eckert,H. and Bajorath,J. (2007) Molecular similarity analysis in virtual screening: foundations, limitations and novel approaches. *Drug Discov. Today*, **12**, 225–233.

Fey,M. and Lenssen,J.E. (2019) Fast Graph Representation Learning with PyTorch Geometric. arXiv:1903.02428.

Gao,Z. *et al.* (2017) Predict drug permeability to blood-brain-barrier from clinical phenotypes: Drug side effects and drug indications. *Bioinformatics*, **33**, 901–908.

Garg,P. *et al.* (2015) Role of breast cancer resistance protein (BCRP) as active efflux transporter on blood-brain barrier (BBB) permeability. *Mol. Divers.*, **19**, 163–172.

Garg,P. and Verma,J. (2006) In Silico Prediction of Blood Brain Barrier Permeability: An Artificial Neural Network Model. *J. Chem. Inf. Model.*, **46**, 289–297.

Kearnes,S. *et al.* (2016) Molecular graph convolutions: moving beyond fingerprints. *J. Comput. Aided. Mol. Des.*, **30**, 595–608.

Kipf,T.N. and Welling,M. (2017) Semi-supervised classification with graph convolutional networks. In, *5th International Conference on Learning Representations, ICLR 2017 - Conference Track Proceedings*. International Conference on Learning Representations, ICLR.

Li,H. *et al.* (2005) Effect of selection of molecular descriptors on the prediction of blood-brain barrier penetrating and nonpenetrating agents by statistical learning methods. *J. Chem. Inf. Model.*, **45**, 1376–1384.

Lingineni,K. *et al.* (2017) The role of multidrug resistance protein (MRP-1) as an active efflux transporter on blood–brain barrier (BBB) permeability. *Mol. Divers.*, **21**, 355–365.

Long,Y. *et al.* (2020) Predicting human microbe–drug associations via graph convolutional network with conditional random field. *Bioinformatics*, **36**, 4918–4927.

Martin,Y.C. *et al.* (2002) Do structurally similar molecules have similar biological activity? *J. Med. Chem.*, **45**, 4350–4358.

Martins,I.F. *et al.* (2012) A Bayesian Approach to in Silico Blood-Brain Barrier Penetration Modeling. *J. Chem. Inf. Model.*, **52**, 1686–1697.

Miao,R. *et al.* (2019) Improved Classification of Blood-Brain-Barrier Drugs Using Deep Learning. *Sci. Rep.*, **9**, 8802.

Moriwaki,H. *et al.* (2018) Mordred: a molecular descriptor calculator. *J. Cheminform.*, **10**, 4.

Morris,M.E. *et al.* (2017) SLC and ABC Transporters: Expression, Localization, and Species Differences at the Blood-Brain and the Blood-Cerebrospinal Fluid Barriers. *AAPS J.*, **19**, 1317–1331.

Muegge,I. and Mukherjee,P. (2016) An overview of molecular fingerprint similarity search in virtual screening. *Expert Opin. Drug Discov.*, **11**, 137–148.

Musgrave,K. *et al.* (2020) PyTorch Metric Learning. arXiv:2008.09164.

Nałęcz,K.A. (2017) Solute Carriers in the Blood–Brain Barier: Safety in Abundance. *Neurochem. Res.*, **42**, 795–809.

Pajouhesh,H. and Lenz,G.R. (2005) Medicinal chemical properties of successful central nervous system drugs. *NeuroRx*, **2**, 541–553.

Palmer,A.M. and Alavijeh,M.S. (2013) Overview of Experimental Models of the Blood-Brain Barrier in CNS Drug Discovery. *Curr. Protoc. Pharmacol.*, **62**, 7.15.1-7.15.30.

Pardridge,W.M. (2012) Drug Transport across the Blood–Brain Barrier. *J. Cereb. Blood Flow Metab.*, **32**, 1959–1972.

Patel,M.M. and Patel,B.M. (2017) Crossing the Blood–Brain Barrier: Recent Advances in Drug Delivery to the Brain. *CNS Drugs*, **31**, 109–133.

Plisson,F. and Piggott,A.M. (2019) Predicting blood–brain barrier permeability of marine-derived kinase inhibitors using ensemble classifiers reveals potential hits for neurodegenerative disorders. *Mar. Drugs*, **17**, 81.

Qi,Q. *et al.* (2021) Stochastic Optimization of Areas Under Precision-Recall Curves with Provable Convergence. arXiv:2104.08736.

Qosa,H. *et al.* (2015) Regulation of ABC efflux transporters at blood-brain barrier in health and neurological disorders. *Brain Res.*, **1628**, 298–316.

Sanchez-Covarrubias,L. *et al.* (2014) Transporters at CNS Barrier Sites: Obstacles or Opportunities for Drug Delivery? *Curr. Pharm. Des.*, **20**, 1422–1449.

Schlichtkrull,M. *et al.* (2018) Modeling Relational Data with Graph Convolutional Networks. In, Gangemi,A. *et al.* (eds), *The Semantic Web. ESWC 2018. Lecture Notes in Computer Science*. Springer, Cham, pp. 593–607.

Schulte-Sasse,R. *et al.* (2021) Integration of multiomics data with graph convolutional networks to identify new cancer genes and their associated molecular mechanisms. *Nat. Mach. Intell.*, **3**, 513–526.

Segarra,M. *et al.* (2021) Blood–Brain Barrier Dynamics to Maintain Brain Homeostasis. *Trends Neurosci.*, **44**, 393–405.

Shaker,B. *et al.* (2021) LightBBB: computational prediction model of blood–brain-barrier penetration based on LightGBM. *Bioinformatics*, **37**, 1135–1139.

Shi,Z. *et al.* (2021) Prediction of blood-brain barrier permeability of compounds by fusing resampling strategies and extreme gradient boosting. *IEEE Access*, **9**, 9557–9566.

Shu,J. *et al.* (2021) Disease gene prediction with privileged information and heteroscedastic dropout. *Bioinformatics*, **37**, i410–i417.

Singh,M. *et al.* (2020) A classification model for blood brain barrier penetration. *J. Mol. Graph. Model.*, **96**, 107516.

Sivandzade,F. and Cucullo,L. (2018) In-vitro blood–brain barrier modeling: A review of modern and fast-advancing technologies. *J. Cereb. Blood Flow Metab.*, **38**, 1667–1681.

Szklarczyk,D. *et al.* (2016) STITCH 5: augmenting protein–chemical interaction networks with tissue and affinity data. *Nucleic Acids Res.*, **44**, D380–D384.

Terstappen,G.C. *et al.* (2021) Strategies for delivering therapeutics across the blood–brain barrier. *Nat. Rev. Drug Discov.*, **20**, 362–383.

Thanapalasingam,T. *et al.* (2021) Relational Graph Convolutional Networks: A Closer Look. arXiv:2107.10015.

Vatansever,S. *et al.* (2021) Artificial intelligence and machine learning-aided drug discovery in central nervous system diseases: State-of-the-arts and future directions. *Med. Res. Rev.*, **41**, 1427–1473.

Wanat,K. (2020) Biological barriers, and the influence of protein binding on the passage of drugs across them. *Mol. Biol. Rep.*, **47**, 3221–3231.

Wang,Z. *et al.* (2018) In Silico Prediction of Blood–Brain Barrier Permeability of Compounds by Machine Learning and Resampling Methods. *ChemMedChem*, **13**, 2189–2201.

Whitfield,A.C. *et al.* (2014) Classics in Chemical Neuroscience: Levodopa. *ACS Chem. Neurosci.*, **5**, 1192–1197.

Wishart,D.S. *et al.* (2018) DrugBank 5.0: a major update to the DrugBank database for 2018. *Nucleic Acids Res.*, **46**, D1074–D1082.

Yu,T. *et al.* (2020) Gradient Surgery for Multi-Task Learning. arXiv:2001.06782.

Yuan,Y. *et al.* (2018) Improved Prediction of Blood–Brain Barrier Permeability Through Machine Learning with Combined Use of Molecular Property-Based Descriptors and Fingerprints. *AAPS J.*, **20**, 54.

Yuan,Y. and Bar-Joseph,Z. (2020) GCNG: graph convolutional networks for inferring gene interaction from spatial transcriptomics data. *Genome Biol.*, **21**, 300.

Zhang,L. *et al.* (2008) QSAR modeling of the blood-brain barrier permeability for diverse organic compounds. *Pharm. Res.*, **25**, 1902–1914.

Zhao,T. *et al.* (2020) Data Augmentation for Graph Neural Networks. arXiv:2006.06830.

Zitnik,M. *et al.* (2018) Modeling polypharmacy side effects with graph convolutional networks. *Bioinformatics*, **34**, i457–i466.